\documentclass[12pt]{iopart}

\renewcommand{\theequation}{\arabic{section}.\arabic{equation}}
\usepackage{amssymb,amsthm,graphicx}
\usepackage{graphicx}
\usepackage{enumerate}
\usepackage{colordvi}

\newcommand{\lhs}{left-hand side}

\newcommand{\R}{\mathbb{R}}
\newcommand{\AAA}{\mathcal{A}}
\newcommand{\BB}{\mathcal{B}}
\newcommand{\DD}{\mathcal{D}}
\newcommand{\OO}{\mathcal{O}}
\newcommand{\eps}{\varepsilon}

\usepackage{color}

\def\*{{\phantom *}}

\begin{document}


\vskip 0.5cm

\title[Non-equilibrium current via geometric scatterers]
{Non-equilibrium current via  geometric scatterers}

\author{Pavel Exner}
\address{Doppler Institute for Mathematical Physics and Applied
Mathematics, \\ Czech Technical University in Prague,
B\v{r}ehov\'{a} 7, 11519 Prague, \\ and  Nuclear Physics Institute
ASCR, 25068 \v{R}e\v{z} near Prague, Czechia} \ead{exner@ujf.cas.cz}

\author{Hagen Neidhardt}
\address{Weierstrass Institute, Mohrenstrasse 39, 10117 Berlin,
Germany} \ead{neidhard@wias-berlin.de}

\author{Milo\v{s} Tater}
\address{Nuclear Physics Institute ASCR, 25068 \v{R}e\v{z} near Prague, Czechia} \ead{tater@ujf.cas.cz}

\author{Valentin A. Zagrebnov}
\address{D\'epartement de Math\'ematiques - Universit\'{e} d'Aix-Marseille (AMU)\\
163 av.de Luminy, 13288 Marseille Cedex 09, and \\
Institut de Math\'{e}matiques de Marseille (UMR 7373)\\
CMI-AMU, Technop\^{o}le Ch\^{a}teau-Gombert \\
39, rue F. Joliot Curie, 13453 Marseille Cedex 13, France } \ead{valentin.zagrebnov@univ-amu.fr}

\begin{abstract}
We investigate non-equilibrium particle transport in the system consisting of a geometric scatterer and two leads coupled to
heat baths with different chemical potentials. We derive expression for the corresponding current the carriers of which are
fermions and analyze numerically its dependence of the model parameters in examples, where the scatterer has a rectangular or
triangular shape.
\end{abstract}

\pacs{03.65.Nk, 72.20.Dp}
\hspace{6em}{\small Mathematics Subject Classification: 81Q35, 82B99} \\
\phantom{AAAAAAAA}{\small Keywords: non-equilibrium steady states, geometric scatterer}
\maketitle


\begin{center}
\emph{Dedicated to the memory of Markus B\"uttiker (1950-2013)}
\end{center}

\tableofcontents
\section{Introduction}

\begin{figure} \label{geomscatt}
 \hspace{8em}\includegraphics[width=5cm]{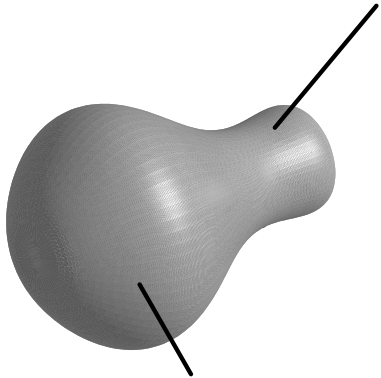}
 \caption{A geometric scatterer}
 \end{figure}

The aim of the present paper is to analyze a stationary fermion transport, giving rise to electric current if the carrier
particles are charged, between two heat baths connected through a geometric scatterer. By the latter we mean a quantum mechanical
system of a mixed dimensionality consisting of a compact two-dimensional manifold  $G$ to which two (infinite) one-dimensional
continuous leads are attached --- cf.~Fig.~\ref{geomscatt} --- the latter are considered as heat and Fermi particle baths at
equilibrium for given temperatures and chemical potentials. This assumption is made
for simplicity. In general, one can think of the leads as of connecting links between the manifold and the baths, however,
since in a one-mode quantum transport the particles are in the asymptotic regime once they leave the manifold, the identification
of the leads with the baths is the easiest way.

The main tool for our analysis is the Landauer-B\"utticker formula, which expresses a steady fermion current (in other words, a
particle flux) through the sample (scatterer) in terms of the transmission probability for the scatterer and of the external
reservoirs equilibrium states. Under  quite general conditions {the formula has been} proved in \cite{AJPP} for the quasi-free
fermions transport in the framework of the $C^{*}$-scattering approach. In fact, this approach allows much more, namely to
construct non-equilibrium steady states and to make contact with non-equilibrium statistical mechanics, see the three-volume
review \cite{AJP} {for a thorough discussion.}

We are not going to prove the Landauer-B\"utticker formula in the present {context}, because for our restricted purpose of study
only the fermion current it can be done repeating \emph{verbatim} the argument used in the analogous situation in \cite{CGZ},
where discrete leads and a discrete sample $G$ were considered. The formula has two essential ingredients: \textit{Fermi functions}
for the thermal statistical distributions of
non-interacting fermions in the leads (reservoirs) and the quantum \textit{transmission probability} between the leads, which
results from the stationary scattering calculations of the single-particle {passage} throughout the sample.

In a sense the problem of a stationary current through the manifold $G$ that we treat here can be regarded as a `continuous'
version of the model discussed in \cite{CGZ} describing a similar effect in the discrete setting. However, the continuous case
has its peculiarities. While the difference is not very important for treating the leads, it  it becomes nontrivial in the
analysis of scattering problem due to its mixed dimensionality, due to which the transport in such systems exhibits unusual
and interesting features. Let us add that the way to describe quantum dynamics of such `strange' scatterers can be traced back
to \cite{ES1}; it is based on construction of admissible Hamiltonians as self-adjoint extensions of a suitable symmetric operator,
starting from the situation when different parts of the configuration space are decoupled.

Let us mention that the \emph{ballistic} conductance --- but not the current --- for a mixed dimensionality scatterer was the
subject of the paper \cite{BGMP} where the model of one-dimensional leads attached to a quantum sphere was investigated. From
the quantum-mechanical point of view, the central problem both in \cite{BGMP} and in the present paper how to match wave-functions
of the leads and the manifold $G$ at the points of their junctions. The mentioned construction based on self-adjoint extension
can be performed in different, equivalent ways, the result being always a family of boundary conditions involving appropriate
generalized boundary values \cite{ES1, BGMP}.

In the present paper we do not reduce ourself to the linear response, i.e. to analysis of the conductance. We study the
non-equilibrium current $I(V, V_g)$ throughout geometric scatterers as a function of {two parameters: the difference:
$V = \mu_2 - \mu_1 \geq 0$ of the two leads \textit{electro-chemical} potentials with the aim} to deduce the quantum
\textit{Ohm law}, {and} the \textit{plunger gate} voltage $V_g$ applied to the scatterer, which controls its \textit{resonant}
quantum conductance. Note that in contrast to \cite{CGZ} {we consider} here {the two `external' parameters \textit{separately}}.
{Of course,} in certain cases {it is} plausible {to work with} the hypothesis $V=V_g$ {which} implies a highly non-linear
current-voltage behaviour \cite{CGZ}, {however,} the current-driving potential difference $V$ and the {gate} voltage $V_g$
controlling the conductance are \textit{a priori} of a different nature. {In particular,} experimentally one can modify the
quantum (resonant) conductance by varying $V_g$, see \cite{CJM}.

{What concerns the literature indicated above,} the ballistic ($V_g = 0$) linear response
$\sigma (\mu, 0):= \partial_{V} I(V, 0)\mid_{V=0}$ with  $\mu = \mu_1 = \mu_2$, was the subject of \cite{BGMP}, whereas the
resonant conductance $\sigma (\mu, V_g)$ modification by {variation of} $V_g$ was considered in \cite{CJM}. {The present paper}
analyses the {current-voltage} dependence $I(V, V_g)$, that is, the Ohm law, parameterized by the plunger gate voltage $V_g$.

Let us briefly review the contents of the paper. In the next section we describe the construction used to couple the wave function
and indicate which of the Hamiltonians obtained in this way might be physically the most relevant. Using the result, we solve
in Sec.~3 the quantum-mechanical scattering problem finding the transmission probability as a function of the involved particle
momenta; we shall also explain how the needed quantities can be computed for a compact manifold $G$. The concluding section is
devoted to discussion of examples in which $G$ is a two-dimensional rectangular and triangular `billiard'. Using the
Landauer-B\"{u}ttiker formula we compute the current and analyze numerically its dependence on the model parameters.

\section{Coupling of leads to the surface}

As we have said, the core of the model is the quantum-mechanical scattering on such a geometric object with leads attached.
The first question to address is how they can be coupled mutually in a way which would conserve the probability current.
Following \cite{ES1} this question was investigated in various papers -- one can mention, e.g., \cite{ES3, Ki, ETV, BGMP}.
Their result cannot be directly used for our purpose, however, since in all those works it was assumed that that the energy
is conserved, i.e. that the particle momentum is the same on both the leads, and have to be modified. One can adopt the (quantum)
transfer matrix approach,
\begin{equation} \label{L}
\left(\begin{array}{c} u(0+) \\ u'(0+) \end{array}\right) \,=\, L\,
\left(\begin{array}{c} u(0-) \\ u'(0-) \end{array}\right)\,.
\end{equation}
with $L$ derived in \cite{ETV} where $u(0\pm)$ and $u'(0\pm)$ are the boundary values of the wave functions on the leads at the
junctions, however, using it as a component of our model, one has to take into account that the particles move with velocities
given by the temperatures and chemical potentials of the reservoirs.

To construct the model indicated in the introduction one has to know how to describe motion of a quantum mechanical particle
on a configuration space of a mixed dimensionality. A general prescriptions how the corresponding self-adjoint Hamiltonians can
be constructed was first formulated in \cite{ES1}. Since the most natural coupling is \emph{local}, we may disregard geometrical
peculiarities of the lead and the surface and illustrate the construction in the setting where a halfline is attached to a plane.
The state Hilbert space is then $\,L^2({\R}_-) \oplus L^2({\R}^2)\,$ and if we neglect physical
constants the Hamiltonian acts on its elements as
\begin{equation*}
\left(\begin{array}{c}\psi_\mathrm{lead} \\ \psi_\mathrm{plane} \end{array}\right) \, \rightarrow \,
\left(\begin{array}{c} -\psi''_\mathrm{lead} \\ -\Delta\psi_\mathrm{plane} \end{array}\right)\,.
\end{equation*}
To make such an operator self--adjoint one has to impose suitable boundary conditions which couple the wave-functions at the
junction.

The boundary values to be used are obvious on the lead side being the columns $\psi_\mathrm{lead}(0-)$ and
$\psi'_\mathrm{lead} (0-)$. On the other hand, in the plane we have to use generalized ones. To understand how to define
them, we note that if we restrict two-dimensional Laplacian to functions vanishing at the origin and take the adjoint to
such an operator, the functions in the corresponding definition domain will have a logarithmic singularity at the origin
\cite{ES1}. The said generalized boundary values $L_j(\psi_\mathrm{plane}),\, \mbox{j=0,1,}\,$ are then given as coefficients
in the corresponding expansion,
 \begin{equation} \label{expans}
\psi_\mathrm{plane}(x) =
L_0(\psi_\mathrm{plane})\,\ln |x| + L_1(\psi_\mathrm{plane}) +
o(|x|)\,,
 \end{equation}
being defined as
 \begin{eqnarray} \label{gen-bv}
 && L_0(\psi_\mathrm{plane})=\lim_{|x|\to
 0}\frac{\psi_\mathrm{plane}(x)}{\ln{|x|}}\,, \nonumber \\[-.5em] \label{gen-bv} \\[-.5em]
 && L_1(\psi_\mathrm{plane})=\lim_{|x|\to 0}\Big[\psi_\mathrm{plane}(x)
 -L_0(\psi_\mathrm{plane}(|x|))\,\ln|x|\,\Big]\,. \nonumber
 \end{eqnarray}
Using these notions we can write the sought boundary conditions as
 \begin{equation} \label{bc}
\begin{array}{rcl}
\psi'_\mathrm{lead} (0+) &=& A\psi_\mathrm{lead} (0+) +
2\pi \bar C L_0(\psi_\mathrm{plane})\,, \\ [.3em] L_1(\psi_\mathrm{plane})
&=& C\psi_\mathrm{lead} (0+) + DL_0(\psi_\mathrm{plane})\,,
\end{array}
 \end{equation}
where $A,D\in\R$ and $C$ is a complex number; it means, in particular, that the coupling depends on four real parameters.
If $C$ is chosen real we get a coupling invariant w.r.t. the time reversal which we will assume throughout in the following.
It is worth noting that the above boundary conditions are generic but do not cover all the self-adjoint couplings leaving out
cases when the coefficient matrix is singular; this flaw can be mended in the standard way \cite{KS} if one replaces (\ref{bc})
by the symmetrized form
 $$
\AAA {\psi_\mathrm{lead}(0+) \choose L_0(\psi_\mathrm{plane})} +
\BB {\psi'_\mathrm{lead}(0+) \choose L_1(\psi_\mathrm{plane})}
=0
 $$
with appropriately chosen matrices $\AAA,\BB$. For our present purpose, however, the generic conditions (\ref{bc}) are sufficient.

The question which boundary conditions are physically `correct' ones is difficult and the general answer to it is not known.
We will not address it here and limit ourselves to mentioning that various choices are used:

\begin{enumerate}[(i)]

\item the simplest possibility is to keep just the term coupling the two parts of the configuration manifold, i.e. to put $A=D=0$.

\item a heuristic way to choose the `natural' coupling was suggested  in \cite{ES3}: comparing the scattering matrix of the
coupling given by (\ref{bc}) with the  low--energy behavior of scattering in the system  of a plane to which  a cylindrical
`tube' is attached, one arrives at the identification
\begin{equation} \label{natcoupl}
A \,=\, {1\over 2\rho}\,,\qquad B = \sqrt{{2\pi\over \rho}}\,, \qquad
C = {1\over\sqrt{2\pi\rho}}\,, \qquad D = -\ln\rho\,,
\end{equation}
where $\rho$ is the contact radius; physical relevance of these conditions was illustrated in \cite{ES3} by explaining the
experimentally observed distribution of resonances in a microwave resonator with a thin antenna.

\item the choice of the coupling amounts to fixing the singularity of the Hamiltonian Green's function at the connection point.
One can do that directly without using boundary conditions explicitly \cite{Ki}.

\end{enumerate}

\noindent What is important is the local character of the coupling which makes it possible to use the above description of the
coupling in any situation where a one-dimensional segment is attached to a smooth two-dimensional surface; this is what we will
use in the following.

\section{Transport through the geometric scatterer}
\setcounter{equation}{0}

\subsection{The transfer matrix}

Equipped with the above notions we can now solve the quantum mechanical part of the problem. The first step consists of finding the
transfer matrix (\ref{L}) for our system. This part is essentially the same as in \cite{ETV} and we include it in order to make
the paper self-contained. The compact manifold $G$ describing the geometric scatterer may or may not have a boundary; we suppose
that the two leads are attached to it at two different interior points $x_1,x_2$ of $G$. The manifold part of the Hamiltonian is
the Laplace-Beltrami operator on the state Hilbert space $L^2(G)$ of the scatterer. It is coupled to the Laplacians on the two
leads by the boundary conditions (\ref{bc}) with the coefficients indexed by $j=1,2$; later we will assume that the couplings are
the same.

The most important object for us is the Green's function $G(.,.;k)$ of the Laplace-Beltrami operator, i.e. the integral kernel of
its resolvent which exists whenever the $k^2$ does not belong to the spectrum. Its actual form depends on the geometry of $G$ but
the diagonal singularity does not. The reason is that the manifold $G$ admits in the vicinity of any point a local Cartesian chart
and the Green's function behaves with respect to those variables as that of Laplacian in the plane,
 \begin{equation} \label{expansG}
G(x,y;k)= -\,\frac{1}{2\pi}\,\ln |x\!-\!y| +\OO(1)\,,
\qquad |x\!-\!y|\to 0\,.
 \end{equation}
Looking for transient solutions to the Schr\"odinger equation, we need a general solution to the Laplace-Beltrami equation on $G$
for the energy $k^2$. Without loss of generality, we may write it as
\begin{equation} \label{surface soln}
u(x) = a_1G(x,x_1;k)+a_2G(x,x_2;k)\,,
\end{equation}
where $x_1,x_2$ are the contact points \cite{Ki}. In view of (\ref{expansG}) the singularities of $u$ at $x_1,x_2$ have the
character (\ref{expans}) and we can evaluate the generalized boundary values (labeled by the point at which they are taken) to be
\begin{equation} \label{gen bv}
L_0[x_j] = -\frac{a_j}{2\pi}\,,\qquad L_1[x_j] =
a_j\xi(x_j,k)+a_{3-j}G(x_1,x_2;k)
\end{equation}
for $j=1,2$, where
\begin{equation} \label{xi1}
\xi(x_j;k) := \lim_{x\to x_j} \left\lbrack G(x,x_j;k)+
\frac{\ln|x\!-\!x_j|}{2\pi} \right\rbrack\,.
\end{equation}
Let $u_j$ be the wavefunction on the $\,j$-th lead. Using the abbreviations $u_j, \,u'_j$ for its boundary values we infer from
the boundary conditions (\ref{bc}) that
\begin{eqnarray*} 
u'_1\,=\, A_1u_1-2\pi \bar C_1a_1 \,, && \qquad
a_1\xi_1+a_2g \,=\, C_1u_1-\,\frac{D_1a_1}{2\pi} \,, \nonumber \\
\nonumber \\
u'_2\,=\, -A_2u_2+2\pi\bar C_2a_2\,, && \qquad
a_2\xi_2+a_1g \,=\, C_2u_2-\,\frac{D_2a_2}{2\pi} \,, \nonumber
\end{eqnarray*}
where we have denoted $g:=G(x_1,x_2;k)$. Note that in the first equation of the second pair we used the opposite sign, because it
is natural to identify the second (i.e., the `right') lead with ${\R}_+$. It is straightforward to rewrite these equations as a
linear system with the unknown $\,u_2,\, u'_2,\, a_1,\, a_2\,$ and to solve it; this gives in particular the transfer matrix,
\begin{equation} \label{geom L}
\hspace{-3em} L = \frac{1}{gC_2}\left(\begin{array}{cc}
C_1Z_2+\frac{A_1}{\bar C_1}\DD & -\frac{\DD}{\bar C_1}
\\ \\
|C_2|^2\left( C_1-Z_1 \frac{A_1}{\bar C_1} \right) -
C_1A_2Z_2-\frac{A_1A_2}{\bar C_1}\DD & \phantom{A}
\frac{A_2}{\bar C_1}\DD + \frac{|C_2|^2Z_1}{\bar C_1}
\end{array}\right)\,,
\end{equation}
where $\,Z_j:= {D_j\over 2\pi} +\xi_j\,$ and $\,\DD := g^2\!-Z_1Z_2\,$; in these formula $\xi_j := \xi(x_j;k)$. It is easy to check
that
\begin{equation}
\det L =
-\,\frac{\overline C_2C_1}{\overline C_1C_2}\,,
\end{equation}
hence $\det L=-1$ as long as we suppose that the coupling is time-reversal invariant and the parameters $C_j$ are real. Note that
the same is true even without this assumption if the couplings are the same. We adopt in the following both hypotheses so that our
model will be characterized by three real parameters $A,C,D$; in that case the transfer matrix (\ref{geom L}) simplifies to the form
\begin{equation}  \label{geom L2}
L \,=\, \frac{1}{g}\left(\begin{array}{cc}
Z_2+\frac{A}{C^2}\DD & -2\frac{\DD}{C^2} \\ \\
C^2-A(Z_1\!+\!Z_2)-\frac{A^2}{C^2}\DD & \phantom{A}
\frac{A}{C^2}\DD +Z_1
\end{array}\right)\,.
\end{equation}

\subsection{Transmission probability}

To make use of the Landauer-B\"uttiker formula we have to know the $S$-matrix, in particular the transmission amplitude via
a quantum gate (dot) for the process in question. To fix direction of the particle current note that according this formula
electrons (fermions\footnote{For definiteness, we think of an electric current between the reservoirs connected through a
hetero\-structure modeled by the manifold $G$, however, the results apply to transport of arbitrary fermions.}) are moving
from the `left' ($x<0$ and $\mu_2$) to the `right' ($x>0$ and $\mu_1$) lead \textit{if} the bias of \textit{electro-chemical}
potentials in these leads is \textit{positive}, i.e. $V = \mu_2 - \mu_1> 0 $.

On the other hand in the nonballistic regime, $V_g \neq 0$, the incoming (`left') and outgoing (`right') particles may have
different momenta $k_1,\, k_2$, hence the formula relating $L$ and $S$ derived in \cite{ETV} is not applicable. It is not
difficult, however, to derive a more general result to replace it. To this aim, we write the scattering solutions to
the `left' and `right' of the manifold $G$ as
\begin{eqnarray}
u_1(x)= \mathrm{e}^{ik_1x}+r\, \mathrm{e}^{-ik_1x} & \quad \dots \quad & x<0  \nonumber \\[-.5em] \label{Ansatz} \\[-.5em]
u_2(x)= t\, \mathrm{e}^{ik_2x} & \quad \dots \quad & x>0 \nonumber
\end{eqnarray}
>From this Ansatz we get the boundary values $u_1^{(\iota)}(0-)$ and $u_2^{(\iota)}(0-)$, $\iota=0,1$, which will enter the
coupling conditions (\ref{bc}) as $\psi_\mathrm{lead}(0)$ and $\psi'_\mathrm{lead}(0)$ for the `left' and `right' lead,
respectively. The relation between $k_1$ and $k_2$ is determined by the conservation of energy, i.e. by the fact that their
squares differ by $V_g$. On the other hand, there is no \emph{a priori} rule relating the energy $k^2$ describing the particle
in the resonator to $k_j^2,\: j=1,2$. We adopt the most simple choice as a \emph{model assumption}
setting
\begin{equation}\label{k12}
{k_j = \sqrt{k^2 -\frac12 (-1)^j V_g} \, , \quad j=1,2 \ \ ,}
\end{equation}
note that such a fixing of the energy scale of the scatterer with respect to those of the leads can be equivalently regarded
as a choice of the `plunger gate voltage' $V_g$ \cite{CJM}. For we the sake of definiteness we suppose that $V_g >0$.

For the moment, however, we keep the three values independent. Inserting the boundary values obtained  from (\ref{Ansatz})
into (\ref{L}) we get the relations
\begin{eqnarray*}
t &=& L_{11} +ik_1L_{12} + r(L_{11}-ik_1L_{12}) \\ [.5em]
ik_2t &=& L_{21} +ik_1L_{22} + r(L_{21}-ik_1L_{22})
\end{eqnarray*}
which represent a system of equations for the reflection and transmission amplitudes being easily solved by
\begin{eqnarray}
r &\!=\!&  -\,\frac{L_{21} + i(k_1L_{22} \!-\! k_2L_{11}) + k_1k_2L_{12}}
{L_{21} - i(k_1L_{22} \!+\! k_2L_{11}) - k_1k_2L_{12}}\,, \nonumber \\[-.5em] \label{rt} \\[-.5em]
t &\!=\!& -\,\frac{2ik_1}{L_{21} - i(k_1L_{22} \!+\! k_2L_{11}) -
k_1k_2L_{12}}\,. \nonumber
\end{eqnarray}
In particular, substituting the elements of the transfer matrix (\ref{geom L2}) into these solutions, we obtain the transmission
probability $t(k_1,k_2,k)$ of our model in the form

 \begin{equation} \label{transm}
\hspace{-4em} -\,\frac{2ik_1g}{C^2-A(Z_1\!+\!Z_2)-\frac{A^2}{C^2}\DD
- ik_1\left(\frac{A}{C^2}\DD +Z_1\right) -ik_2\left(Z_2+\frac{A}{C^2}\DD \right) +
2k_1k_2 \frac{\DD}{C^2}}\,,
 \end{equation}
the quantities $g,\,Z_j,\,\DD$ being functions of $k$. Needless to say, we consider only the situation when the leads
are not decoupled from the resonator, $C\ne 0$, in which case the expression in the denominator makes sense.

The behaviour of the function $t$ is in general quite complex. It has been analyzed in previously in a particular situation
when $k_1=k_2=k$ (i.e., in the ballistic regime, $V_g =0$) and $G$ is a sphere to which the leads are coupled in polar or
non-polar positions and in particular ways \cite{Ki, ETV, BGMP}. Such systems have numerous resonances corresponding to
energies for which the terms linear in the momentum dominate in the denominator, while away from them the transmission is
governed by the quadratic term and decreases with increasing energy. It is expected that the same will be true for a much
wider class of geometric scatterers.

\subsection{The resonator quantities}

To make use of the above results one must be able to evaluate the functions $g,\,Z_j,\,\DD$ entering the formula (\ref{transm}).
Since $G$ is supposed to be compact, so the Laplace-Beltrami operator on it has a purely discrete spectrum, one can use the
corresponding spectral analysis; we recall the procedure following again essentially the discussion in \cite{ETV}.
The eigenvalues $\{\lambda_n\}_{n=1}^{\infty}\,$ number in the ascending order with the multiplicity taken into account
correspond to eigenfunctions $\{\phi_n\}_{n=1}^{\infty}\,$ which form an orthonormal basis in $L^2(G)$. The common Green's
function expression then gives
\begin{equation} \label{g}
g(k)\,=\,\sum_{n=1}^{\infty}\, \frac{\phi_n(x_1)
\overline{\phi_n(x_2)}}{\lambda_n\!-k^2}\,.
\end{equation}
To express the remaining three values. $Z_1,\, Z_1$  and  $\DD$, we have to compute the regularized limit (\ref{xi1}).
Expanding the logarithm into the Taylor series, we can rewrite the sublimit expression as
$$
G(x_j+\sqrt{\eps}n,x_j;k)+\frac{\ln{\sqrt\eps}}{2 \pi}
\,=\, \sum_{n=1}^\infty\left(\frac{\phi_n(x_j+\sqrt\eps
n)\phi_n(x_j)}{\lambda_n-k^2}-\frac{(1\!-\!\eps)^n}{4\pi
n}\right)\,,
$$
where $n$ is a unit vector in the local chart around the point $x_j$. Unfortunately, interchanging the limit with the sum
is not without risk since the series does not converge uniformly; to see that the result may indeed depend on the regularization
procedure, it is sufficient to replace $\sqrt\eps$ by $c\sqrt\eps$ at the \lhs. To gauge the possible non-uniqueness, let us
compute the difference
$$
\xi(x_j,k) - \xi(x_j,k') \,=\, \lim_{\eps\to
0+}\sum_{n=1}^{\infty}\left(\frac{\phi_n(x_j+\sqrt{\eps
n})\overline{\phi_n(x_j)}}{\lambda_n\!-k^2} \,-\,\frac{\phi_n(x_j
+\sqrt{\eps n})\overline{\phi_n(x_j)}}{\lambda_n-k'^2}\right)\,.
$$
This sum is already uniformly convergent, because by standard semiclassical estimates \cite[XIII.16]{RS} the sequence
$\,\{\| \phi_n\|_\infty\}_{n=1}^{\infty}\,$ is bounded under our assumptions and $\,\lambda_n= 4\pi|G|^{-1}n +\OO(1)\,$ as
$\,n\to\infty\,$, hence
$$
\frac{1}{\lambda_n\!-k^2}\,-\,\frac{1}{\lambda_n\!-k^{'2}}
\,\sim\,\frac{1}{n^2}\,,
$$
and therefore
\begin{equation} \label{diff1}
\xi(x_j,k) - \xi(x_j,k') \,=\,
\sum_{n=1}^{\infty}\,\left(\frac{|\phi_n(x_j)|^2}{\lambda_n\!-k^2}
\,-\,\frac{|\phi_n(x_j)|^2}{\lambda_n\!-k^{2'}}\right)\,.
\end{equation}
>From the same reason one can claim that
\begin{equation}
\tilde\xi (x_j,k)\,: =\,
\sum_{n=1}^{\infty}\,\left(\frac{|\phi_n(x_j)|^2}
{\lambda_n\!-k^2}\,-\,\frac{1}{4\pi n}\right)
\end{equation}
makes sense and $\,\xi (x_j,k)- \tilde\xi(x_j,k)\,$ is independent of $\,k\,$. We have therefore
\begin{equation} \label{xi2}
\xi (x_j,k) \,=\, \sum_{n=1}^{\infty}\, \left(\frac{|\phi_n(x_j)|^2}
{\lambda_n\!-k^2}\,-\,\frac{1}{4\pi n}\right) \,+\, c(G)\,.
\end{equation}
The constant $c(G)$ depends on the manifold $G$ only and we may neglect it unless a particular coupling has to be fixed,
because a nonzero value of $c(G)$ amounts just to a coupling renormalization: $\,D_j$ has to be changed to $\,D_j\!+2\pi c(G)\,$.
Little is known about a proper choice of $c(G)$, we can only recall that for a flat rectangular $G$ treated in \cite{ES3} an
agreement with the experiment was found using $c(G)=0$.

\medskip

\noindent \textbf{Remark:} Note that there is another way to switch on the plunger gate potential directly on the resonator,
see \cite{CJM}. It is experimentally realizable as a shift of the resonator spectrum by the gate voltage $\widehat{V}_g$:
$\{\lambda_n \rightarrow \widehat{\lambda}_n\}_{n=1}^{\infty}\,$, where $\widehat{\lambda}_n := {\lambda}_n + \widehat{V}_g$.
To include this kind of the gate potential into our scheme we have only to modify correspondingly the Green function
$G \rightarrow \widehat{G} \ $ by the spectral shift $\widehat{V}_g$ and to recalculate the coefficients in representation
(\ref{transm}).

\section{Examples}
\setcounter{equation}{0}

We stress that the voltage difference appearing in (\ref{k12}) in itself does not produce any particle current although
the $S$-matrix is in general nontrivial. It is the lack of equilibrium between fermions in the two leads (playing role of
external heat baths). It is true even in the ballistic regime when we consider an infinitesimal difference between the two
equilibria. As mentioned in the introduction, the non-equilibrium stationary flux of particles is expressed by the
Landauer-B\"{u}ttiker formula, which involves the quantum mechanical transmission probability and two non-equal
Fermi-Dirac functions for the left and right lead --- see, e.g., \cite{CGZ}.

Our main aim now is to find the conductivity of the geometric scatterer for different values of the electro-chemical
potentials $\mu_1 , \mu_2$ and/or different temperatures. The quantity of interest is the stationary current $I$ between
the two leads,
\begin{equation} \label{xi2}
I = 2\pi
\int_{V_{g}/2}^\infty [f_{\beta}(\lambda-\mu_2)-f_{\beta}(\lambda-\mu_1)]\,
|t(k_1,k_2,k)|^2\, \mathrm{d} \lambda\,,
\end{equation}
where $\lambda$ is related to the sample energy by $k=\sqrt{\lambda}$ and the momenta $k_j$ are given by (\ref{k12}), and
furthermore, $f$ is the Fermi-Dirac function,
\begin{equation}
f_{\beta}(\lambda)\,:=\, \frac{1}{e^{\beta\lambda}+1}
\end{equation}
corresponding to the temperature $\beta^{-1}$. The conductivity $\sigma$ is obtained as derivative of $I$ with respect to
the corresponding potential bias $V = \mu_2-\mu_1 >0 $. In particular, if the potential difference is infinitesimal (the so-called
\emph{linear-response regime}), we obtain for the conductivity the usual Landauer-B\"{u}ttiker formula in which $|t(k_1,k_2,k)|^2$
is integrated over $- \partial_{\lambda} f(\lambda - \mu)$, where $\mu = \mu_1=\mu_2$ as, for instance, in \cite{BGMP} for the
ballistic regime $k=k_1=k_2$. Since spherical scatterers were analyzed in this paper as well as in \cite{ETV}, we shall treat
in the following two other examples.


\subsection{A rectangular resonator} \label{rectres}

In the first one $G$ is a rectangle with Dirichlet boundary. We note first that numerical treatment of (\ref{transm}) requires
certain caution, cf. the discussion in \cite{CJM} for the discrete case. All the three quantities $g,\,Z_j,\,\DD$ entering the
 formula (\ref{transm}) are infinite series depending on $k$ whose terms are indexed by a pair of indices (the formal index
 $n$ is a pair $n=(n_x,n_y)$). It is useful to limit the maximal eigenvalue $\lambda_\mathrm{max}$ first, then find all levels
 below this value and sum over all the corresponding pairs of indices.

We use the standard eigenfunctions and eigenvalues \cite{ES3} to compute (\ref{g}) and (\ref{xi2}) (putting $c(G) = 0$ following
\cite{ES3} as mentioned above), to be inserted into (\ref{transm}). If the rectangle is $[0,c_1]\times[0,c_2]$, then
$$
\psi_{n_x,n_y}(x,y)=\frac{2}{\sqrt{c_1c_2}}
\sin\left(n_x\frac{\pi}{c_1}x\right)
\sin\left(n_y\frac{\pi}{c_2}y\right)
$$
and
$$
\lambda_{n_x,n_y}=\left(\frac{n_x\pi}{c_1}\right)^2+
\left(\frac{n_y\pi}{c_2}\right)^2
$$
are the eigenfunctions and eigenvalues, respectively.

The series converge slowly. In order to achieve three-digit precision we had to sum up typically $5.10^6$ terms. We present our
results on Figs.~2 and~3.  

We chose $c_1=2, c_2=1$ and three positions where the leads are attached as follows: the incoming one is always attached at
$x_1=0.2, y_1=0.1$ and the three outgoing positions are $x_2=1.8, y_2=0.9$, or $x_2=0.2, y_2=0.9$, or $x_2=1, y_2=0.5$,
respectively. Fig.~2 shows dependence of transmission probability $|t|^2$ on the energy $\lambda$ of the particle in the
resonator for the three indicated positions of outgoing leads. We see that some resonances appear at the same energy values,
however, their significance changes being influenced, in particular, by the value of the eigenfunctions, which generates them,
at the junction point.
\begin{figure}
\begin{center}
\includegraphics[angle=0,width=1.\textwidth]{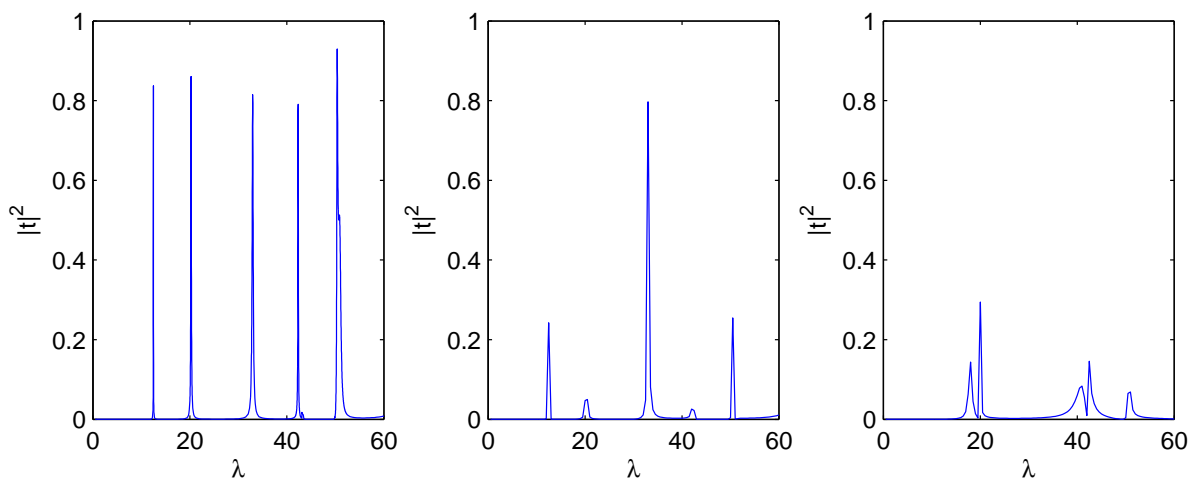}
\caption{Dependence of $|t|^2$ on $\lambda=k^2$ for three positions of
outcoming leads, namely $x_2=1.8,\, y_2=0.9$, or $x_2=0.2,\, y_2=0.9$, or $x_2=1,\,
y_2=0.5$, respectively. The incoming lead is always attached at $x_1=0.2,\,
y_1=0.1$.}
\end{center}
\end{figure}

Note that the transmission probability coefficient $|t|^2$ may in general exceed unity. It clear from (\ref{Ansatz}) and
(\ref{k12}) that the necessary condition for this to happen is $V_g>0$ because the scattering unitarity requires, in particular,
that $|r|^2 + \big(\frac{k_2}{k_1}\big)^2 |t|^2 =1$. Knowing the transmission probability we can proceed to compute the current.
As this part does not differ substantially for different shapes of the scatterer, we will work it out only in the situation of
our following example.


\subsection{A triangular resonator}

In our next example $G$ has a triangular shape. There are three triangular figures with Dirichlet boundary conditions for
which the Laplacian eigenfrequencies and eigenfunctions are expressible via elementary transcendental functions \cite{Ju}. We
choose the triangle with vertices $v_x^{(1)}=0, v_y^{(1)}=0$, $v_x^{(2)}=0, v_y^{(2)}=4\sqrt{3}$, $v_x^{(3)}=3, v_y^{(3)}=\sqrt{3}$.
The eigenfrequencies are labeled by two positive integers $k,n$ with $n>k$, either both even or both odd,
$$
\lambda_{n,k}=\frac{\pi^2}{108}(k^2+3n^2)\,.
$$
All the eigenfrequencies are simple. The corresponding eigenfunctions (normalized to unity) read
\begin{equation}
\begin{array}{l c l}
\psi_{k,n}(x,y)&=&\frac{\sqrt{2}}{3 \sqrt[4]{3}} \left(\sin
\left(\frac{\pi n x}{6}\right)\sin \left(\frac{\pi k \left(y+2
\sqrt{3}\right)}{6 \sqrt{3}}\right) \right. + \nonumber\\ \\
 & &\sin \left( \frac{\pi n \left(\sqrt{3} x-3 y\right)}{12
\sqrt{3}} \right) \sin \left(\frac{\pi k \left(\sqrt{3} x+y-4
\sqrt{3}\right)}{12 \sqrt{3}}\right)- \nonumber\\ \\
 & &\left. \cos \left(\frac{\pi \left(\sqrt{3} n x+3 n y-6
\sqrt{3}\right)}{12 \sqrt{3}}\right) \sin \left(\frac{\pi k
\left(\sqrt{3} x-y+4 \sqrt{3}\right)}{12 \sqrt{3}} \right) \right).
\nonumber\\
\end{array}
\end{equation}

Next we have to choose points where to attach the leads. There are some interesting combinations, in particular,
\begin{enumerate}
\item{near the vertex $v^{(1)}$, $\:x_1=0.1,\, y_1=0.2$, and in the center of
mass, $\,x_2=1,\, y_2=5/\sqrt{3}$}
\item{from the maximum of the
ground state $x_1=1.195408,\, y_1=2.392313$ to the maximum of the
first excited state $x_2=1.142144,\, y_2=1.645060$, i.e. between two
points in central area}
\item{between two most distant vertices $v^{(1)}$ and $v^{(2)}$, $\:x_1=0.1,\, y_1=0.2$ and $x_2=0.1$,\, $y_2=6.6$.}
\end{enumerate}

\begin{figure}
\begin{center}
\includegraphics[angle=270,width=1.\textwidth]{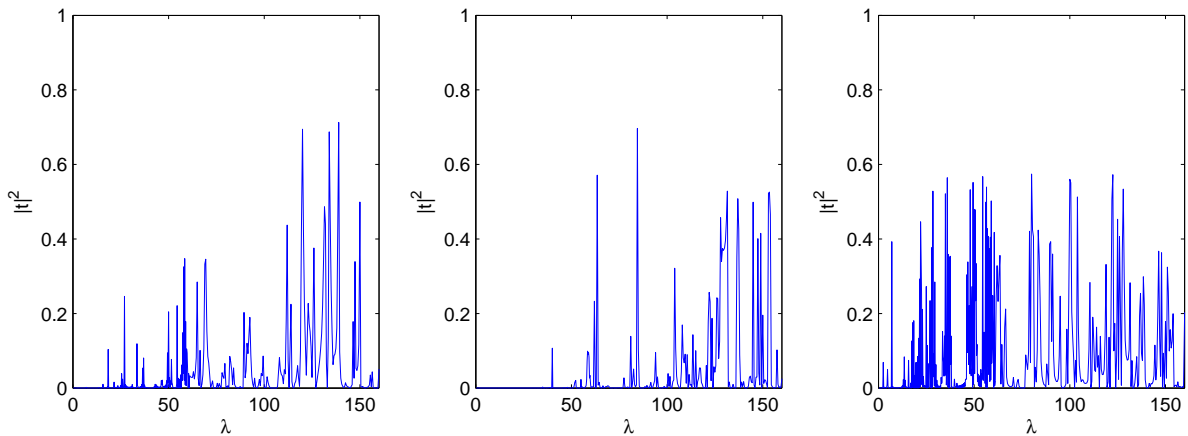}
\caption{Dependence of $|t|^2$ on $\lambda$ and on the points, where the leads are attached. The positions are described
in the text below formula (4.3).}
\end{center}
\end{figure}


\subsection{Stationary current}

Let us now compute the stationary current $I=I(\mu_1,\mu_2,V_g)$ for the triangular resonator and inspect its dependence on
the parameters of the model. It is clear from (\ref{xi2}) it can be quite intricate and several factors play role, in particular,
the spectral properties of the resonator together with the plunger-gate voltage $V_g$ and junction positions, all those determining
the transmission probability $|t(k_1,k_2,k)|^2$, in combination with the smearing coming from the shifted Fermi-Dirac functions
$f_{\beta}(\lambda-\mu_2)-f_{\beta}(\lambda-\mu_1)$. It is clear that the latter approaches the characteristic function of the
interval $(\mu_1, \mu_2)$ in the zero-temperature limit, $\beta\to\infty$.

\begin{figure}
\begin{center}
\includegraphics[width=8cm]{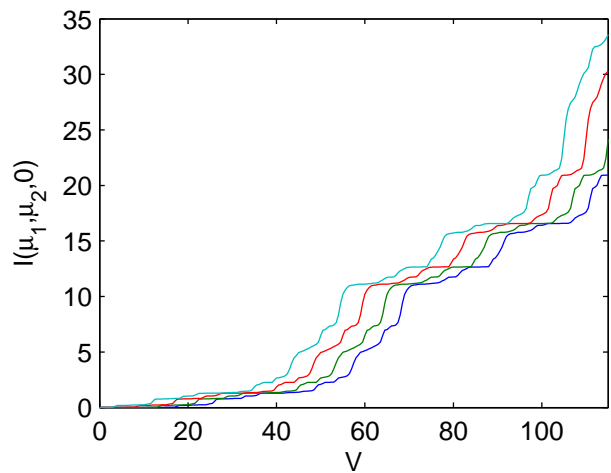}
\caption{Variation of the current $I_\infty$ in the ballistic regime, $V_g=0$,  as a function of the bias $V=\mu_2-\mu_1$. Four
choices are shown, $\mu_1=1,5,10,15$, with $\beta=25$, and the plot moves to the \emph{left} as $\mu_1$ increases; in the
colour-online version they are represented by curves in blue, green, red, and cyan, respectively.}
\end{center} \label{Fig4}
\end{figure}

\begin{figure}
\begin{center}
\includegraphics[width=1.\textwidth]{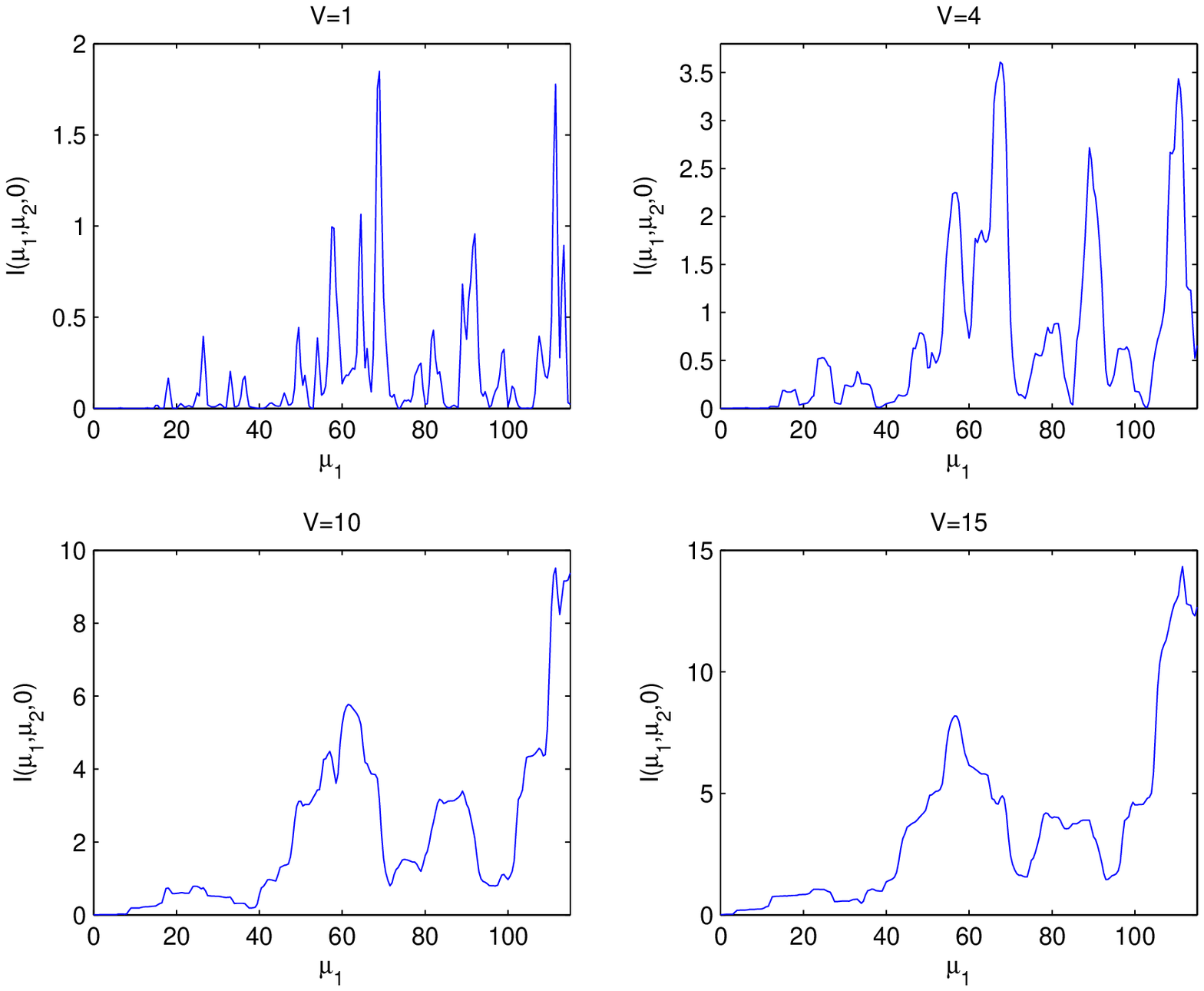}
\caption{Variation of the current $I_\infty$ in the ballistic regime, $V_g=0$, as a function of $\mu_1$, while the potential bias
$V=\mu_2-\mu_1$ is fixed; we present four choices of $V=1,4,10,15$, with $\beta=25$.}
\end{center} \label{Fig5}
\end{figure}

For the sake of definiteness we always consider the situation (i) considered above, i.e. one lead attached near the vertex
$v^{(1)}$, $\:x_1=0.1,\, y_1=0.2$, and the other in the center of mass, $\,x_2=1,\, y_2=5/\sqrt{3}$. We start with the ballistic
regime, $V_g=0$. In the next Figure~4 we plot the current $I=I(\mu_1,\mu_2,0)$as a function of the potential bias $V=\mu_2-\mu_1$
for four different choices of $\mu_1$ and the inverse temperature $\beta=25$. We see that $I(\mu_1,\mu_1+V,0)$ as a function of
$V$ is monotonous increasing. This fact can be easily understood looking at the shape of the function
$f_{\beta}(\lambda-\mu_1-V) -f_{\beta}(\lambda-\mu_1)$ which smears the spectral peaks of $|t(k_1,k_2,k)|^2$; at the considered
high low temperature it is roughly a box the width of which increases with $V$; the curve keeps roughly its shape and moves to
the left as $\mu_1$ increases.

Considering the ballistic regime again, we illustrate in Figure~5
variation of the current $I(\mu_1,\mu_2=\mu_1+V,0)$
as a function of the electrochemical potential $\mu_1$ in the left reservoir for several values of the potential bias
$V=\mu_2-\mu_1$. The plot is again easily understood taking into account the almost-box-shape of the function
$f_{\beta}(\lambda-\mu_1-V) -f_{\beta}(\lambda-\mu_1)$: its with width increases with $V$, and as a consequence the resonance
peak structure becomes washed out for large bias. Another effect to notice concerns the behaviour for small $\mu_1$. If the bias
is small, the current is negligible there, since the transmission probability $|t(k_1,k_2,k)|^2$ is tiny only for the spectral
parameter $\lambda\lesssim 15$; it becomes visible at larger values of $V$ due to the `Fermi-Dirac averaging'.

\begin{figure}
\begin{center}
\includegraphics[width=1.\textwidth]{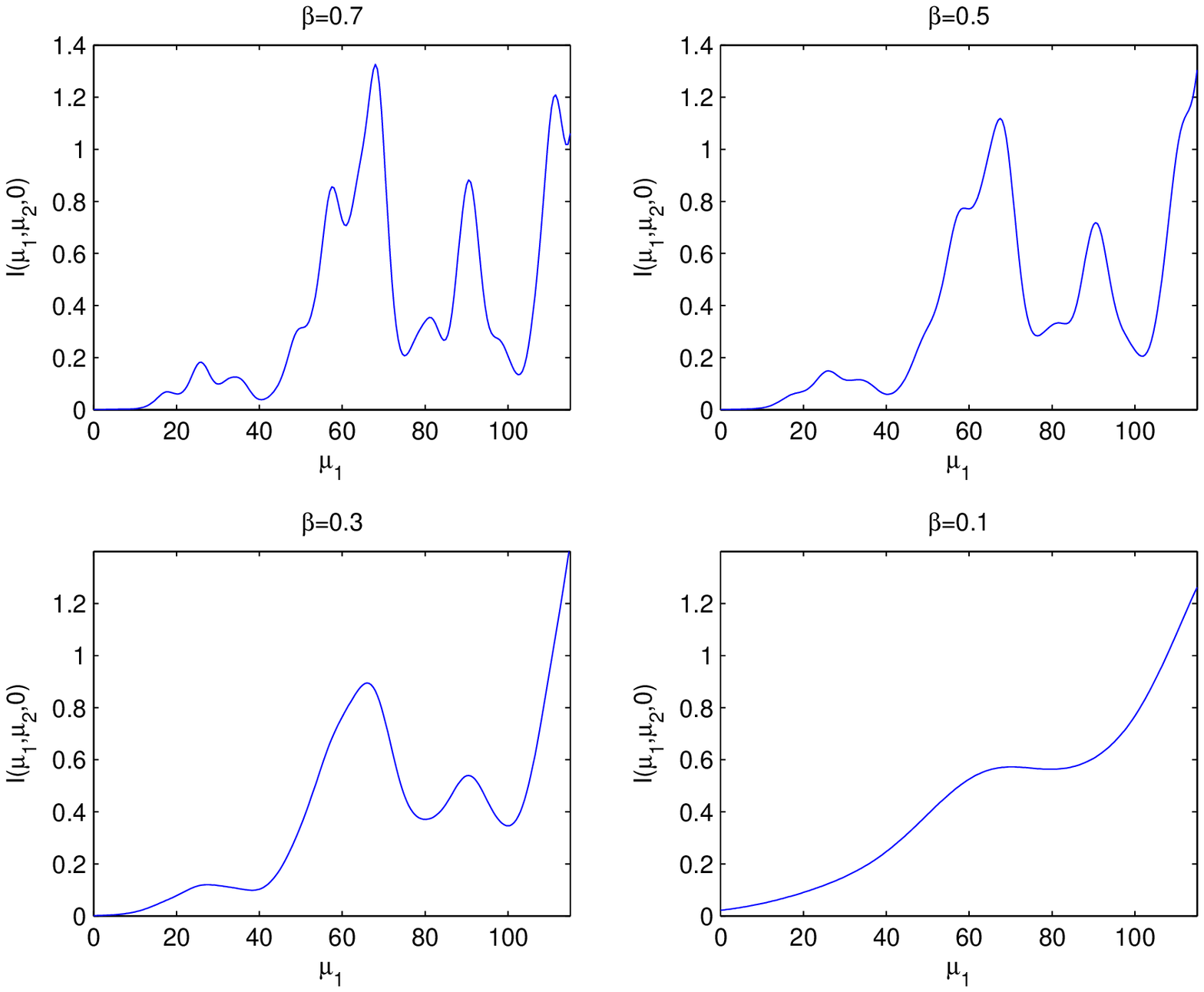}
\caption{The current $I_\infty$ as a function of $\mu_1$ in the ballistic regime for $V=2$ and a sequence of increasing
temperatures corresponding to $\beta=0.7,\,0.5,\,0.3,\,0.1$.}
\end{center} \label{Fig6}
\end{figure}

Still in the ballistic regime, Figure~6
treats the same situation as before, now illustrating the dependence of the
current on the temperature. We choose a small potential bias, $V=2$. In the low-temperature regime the transmission probability
$|t(k_1,k_2,k)|^2$ is then integrated with a function close to a narrow box which produces a plot `interpolating' between the
first two graphs of Figure~\ref{Fig5}. If the temperature is considerably higher,
$f_{\beta}(\lambda-\mu_1-V) -f_{\beta}(\lambda-\mu_1)$ turns into a widely spread smooth peak leading again to washing out the
resonance structure.

\begin{figure}
\begin{center}
\includegraphics[width=1.\textwidth]{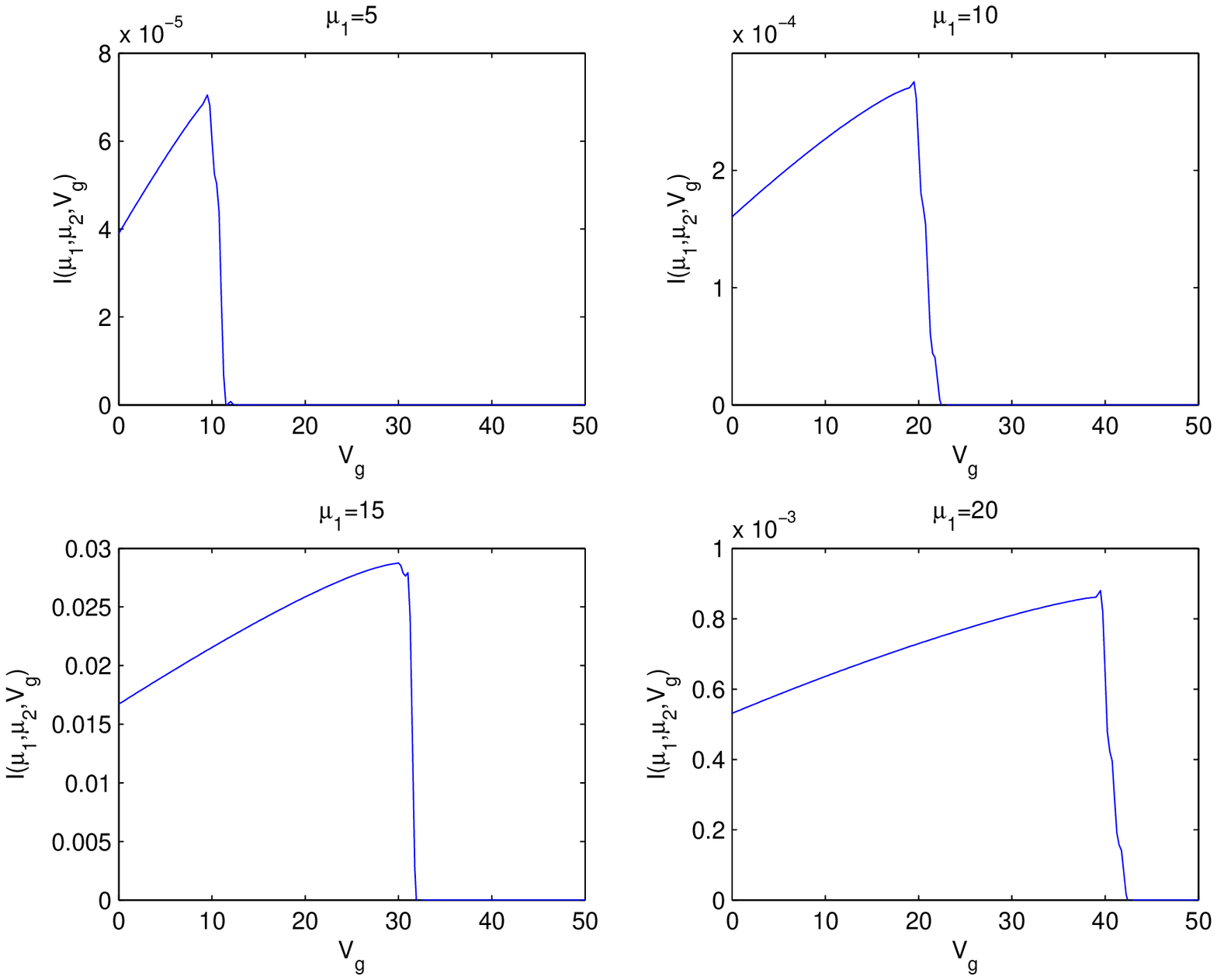}
\caption{Variation of the current $I_\infty$ in the non-ballistic regime on the plunger-gate voltage $V_g$ for fixed $V=1$ and
different values of $\mu_1=5,10,15,20$, with $\beta=25$.}
\end{center} \label{Fig7}
\end{figure}

Turning to the non-ballistic regime, we illustrate in the next figures the dependence of the current $I(\mu_1,\mu_2=\mu_1+V, V_g)$
on the plunger-gate potential $V_g$, according to our convention considered nonnegative. We choose again a small potential bias,
$V=1$, and four different values of the electrochemical potential $\mu_1$. For zero temperature the only contribution to the
current comes in view of (\ref{xi2}) from the values of the spectral parameter $\lambda$ belonging to the interval $(\mu_1, \mu_2)$,
and it follows from assumption (\ref{k12}) that the current vanishes unless $V_g \leq 2\mu_2$. This no longer true for positive
temperatures, however, in the low temperature regime such as $\beta=25$ considered in Figure~7
the current above the value $2\mu_2$ is still negligible. This is easy to understand taking into account the exponential fall-off
of the smearing function. Figure~7
also shows that below this threshold value the current increases. This effect comes from the increase
of the transmission probability $|t(k_1,k_2,k)|^2$ with the of the plunger-gate potential $V_g$. We have remarked at the end of
Sec.~\ref{rectres} that this quantity may exceed one for $V_g>0$, and without showing the graph we claim that it indeed happens
here for $V_g$ large enough.

\begin{figure}
\begin{center}
\includegraphics[width=1.\textwidth]{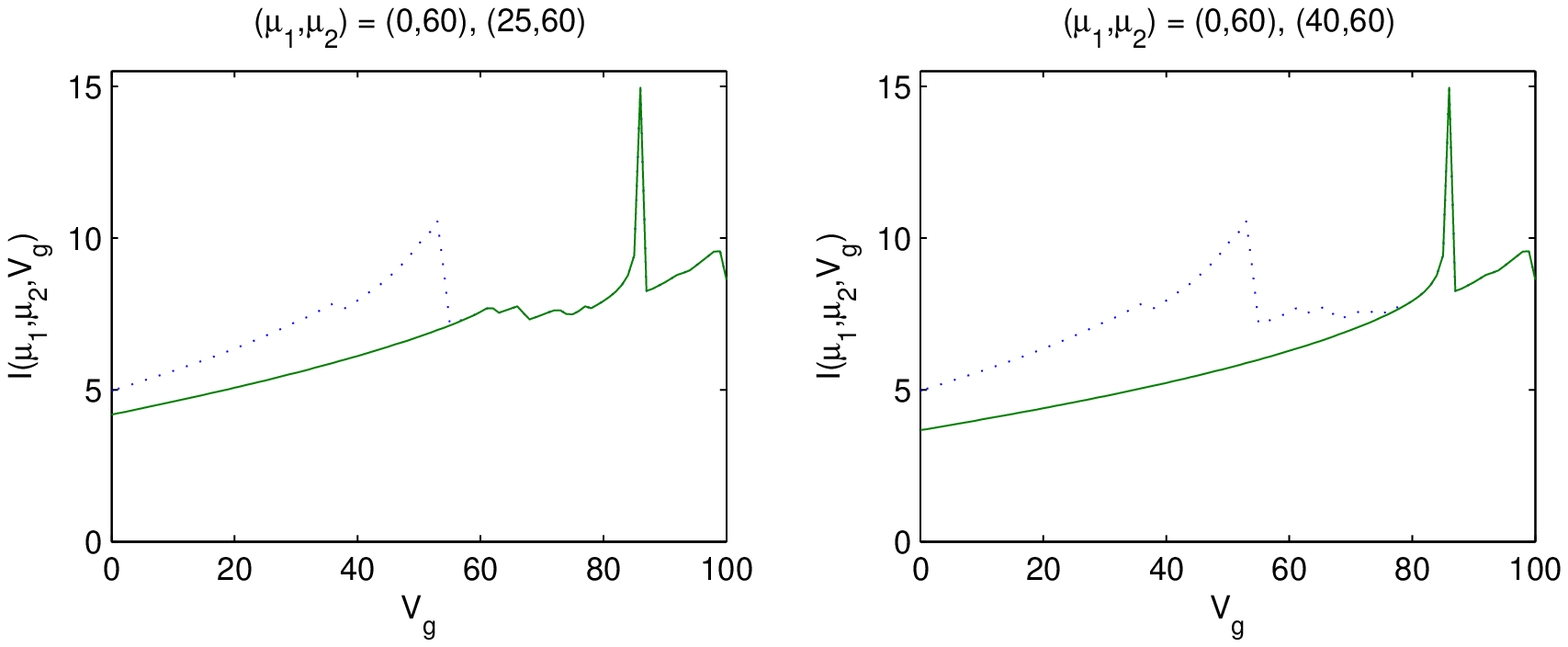}
\caption{Change of the current due to variation of $V$, with $\beta=25$. The solid curve refers to $\mu_1=0,\: \mu_2=60$, the
dotted ones to two other values of $\mu_1$.}
\end{center} \label{Fig8}
\end{figure}

However, the increase is in general not monotonous, because the resonance effects are not complete suppressed by the smearing with
$f_{\beta}(\lambda-\mu_1-V) -f_{\beta}(\lambda-\mu_1)$. This is shown on our last picture, Figure~8,
where we plot again the current \emph{vs.} the plunger-gate voltage. Now we stay below the `threshold' value $2\mu_2$ and we see
that the plot may have peaks depending on both values of the electrochemical potentials. Furthermore, we illustrate how the plot
changes with $\mu_1$. At both graphs the solid curve shows the situation for $\mu_1=0$ and $\mu_2=60$, the dotted ones show what
happens if $\mu_1$  changes to 25 and 40, respectively. We see that the upper part
of the plot changes only a little, while for smaller values of $V_g$ the change of $\mu_1$ reveals an additional structure.

\subsection*{Acknowledgments}

The research was supported by the Czech Science Foundation within the project 14-06818S. VAZ is thankful to Valeriu
Moldoveanu for instructive discussions concerning, in particular, the numeric implementation of the non-equilibrium current
analysis in the case of discrete geometric scatterers. The authors also acknowledge hospitality extended to them during the
respective visits: VAZ and HN to the Nuclear Physics Institute ASCR, the stay which triggered this paper, and PE to Centre
de Physique Th\'eorique, CNRS, Marseille-Luminy.


\end{document}